\documentclass[10pt,aps,prd,floatfix,showpacs,twocolumn,superscriptaddress,nofootinbib]{revtex4-2}

\usepackage{graphicx}
\usepackage{amsmath, amsfonts, amssymb, bm,color}

\usepackage{amstext}
\usepackage{mathrsfs}
\usepackage{cleveref}
\usepackage{slashed}

\newcommand{\corr}[1]{{\color{black}#1}}

\newcommand{\pder}[2]{\frac{\partial #1}{\partial #2}}
\newcommand{\der}[2]{\frac{d #1}{d #2}}
\newcommand{\inv}[1]{\frac{1}{#1}}

\begin{document}
	
	\title{Proportionality of gravitational and electromagnetic radiation by an electron in an intense plane wave}
	\author{G. Audagnotto}
	\email{audag@mpi-hd.mpg.de}
	\author{C. H. Keitel}
	\author{A. Di Piazza}
	\email{dipiazza@mpi-hd.mpg.de}
	\affiliation{Max Planck Institute for Nuclear Physics, Saupfercheckweg 1, D-69117 Heidelberg, Germany} 
	\begin{abstract}
		Accelerated charges emit both electromagnetic and gravitational radiation. Classically, it was found that the electromagnetic energy spectrum radiated by an electron in a monochromatic plane wave is proportional to the corresponding gravitational one. Quantum mechanically, it was shown that the amplitudes of graviton photoproduction and Compton scattering are proportional to each other at tree level. Here, by combining strong-field QED and quantum gravity, we demonstrate that the amplitude of nonlinear graviton photoproduction in an arbitrary plane wave is proportional to the corresponding amplitude of nonlinear Compton scattering. Also, introducing classical amplitudes we prove that the proportionality relies on the semiclassical nature of the electron's motion in a plane wave and on energy-momentum conservation laws, leading to the same proportionality constant in the classical and quantum case. These results deepen the intertwine between gravity and electromagnetism into both a nonlinear and a quantum level.
	\end{abstract}
	
	\maketitle
	
	\section{Introduction}

	Recently large interest has awakened in the connection between gravity and Standard Model gauge theories. There are multiple motivations driving this research area, concerning both fundamental and technical aspects \cite{Bern_2002, Borsten_2020, Bjerrum-Bohr_2014, Bern_2008}. It is known that a satisfactory description of quantum gravity is not available today, nevertheless the low-energy limit of any possible model should be properly connected to the Standard Model and the classical theory of gravity, which are both experimentally well verified \cite{Peskin_B_1995,Turyshev_2008}. The fact that canonical quantum gravity is not strictly renormalizable does not affect these considerations and leading-order calculations are of notable interest. Indeed, it is possible to describe the dynamics of massive objects through the classical limit of scattering amplitudes \cite{Bautista_2019}. By taking into account the proper Feynman diagrams, one can find corrections to the Coulomb or Newton potential \cite{Bjerrum-Bohr_2002, Bjerrum-Bohr_2018}.
	
	The relation between gravity and gauge theories manifests itself, for instance, through the Kawai-Lewellen-Tye relations \cite{Kawai_1985, Bern_2008} derived in the context of string theory, which relate graviton and gauge-bosons tree-level amplitudes. These relations suggest a more fundamental connection between general relativity and gauge theories: at the semiclassical level gravity actually behaves as a double copy of a gauge theory \cite{Bern_2002,Bern_2008,Bern_2010} (see Refs. \cite{Adamo_2017, Adamo_2020b} for studies about double copy in a background plane wave electromagnetic field). Moreover, general considerations about conservation laws \cite{Goebel_1980, Choi_1993,Choi_1994} in tree-level diagrams have been exploited to relate, for example, graviton photoproduction and QED Compton scattering. These factorization properties and various other techniques \cite{Elvang_2015} played a central role in calculating on-shell amplitudes involving gravitons as Compton-like scattering and photoproduction \cite{Holstein_2006_a,Bjerrum-Bohr_2014,Bjerrum-Bohr_2015,Holstein_2006_b,Ahmadiniaz_2016,Ahmadiniaz_2019_a,Bastianelli_2012}. 
	
	On a different side, the recent detection of gravitational waves has attracted a lot of attention \cite{LIGOScientific_2016}. Despite the outstanding experimental result, the lack of measurable events makes it necessary to search for different sources of these perturbations. The classical interplay between gravity and electromagnetism has a long history and in this context Refs. \cite{Nikishov_1989,Nikishov_2010} are of particular interest. In these works it is proved that a proportionality exists between the electromagnetic and gravitational energy spectra of a charge driven by a monochromatic plane wave.
	
	The relations studied in Refs. \cite{Goebel_1980, Choi_1993,Choi_1994} lead to the following result for the linear graviton photoproduction by an electron of charge $e<0$ and mass $m$
	\begin{equation}
		\label{M_g_M_gamma}
		\varepsilon_{i,\alpha} \varepsilon^*_{f,\mu}\varepsilon^*_{f,\nu} M_{e\gamma \rightarrow e g}^{\alpha \mu\nu}   = H \varepsilon_{i,\alpha} \varepsilon^*_{f,\mu} M^{\alpha\mu}_{e\gamma \rightarrow e \gamma}
	\end{equation}
	where
	\begin{equation}\label{H constant}
		\begin{split}
			H = & -\frac{\kappa}{2 e}
			\left(
			\frac{p_i \cdot \varepsilon^{*}_f k_f \cdot p_f - p_f \cdot \varepsilon^{*}_f  	k_f \cdot  p_i   }{k_i \cdot k_f} 
			\right),
		\end{split} 
	\end{equation}
	with $\kappa = \sqrt{32 \pi G}$ in units where $\hbar=c=1$ and $\alpha=e^2/4\pi\approx 1/137$, which are used throughout. Here, the left-hand-side is the amplitude associated to the graviton photoproduction $ e(p_i) + \gamma(k_i, \varepsilon_i)\rightarrow e(p_f) + g(k_f, \varepsilon_f \varepsilon_f) $ and the right-hand side the linear Compton scattering $ e(p_i) + \gamma(k_i, \varepsilon_i)\rightarrow e(p_f) + \gamma(k_f, \varepsilon_f) $. Note that the polarization tensor $\varepsilon_f^{\mu\nu}$ of the graviton is assumed to be written as $\varepsilon_f^{\mu\nu}=\varepsilon_f^{\mu}\varepsilon_f^{\nu}$, where $\varepsilon_f^{\mu}$ is the helicity polarization four-vector of a photon with the same four-momentum \cite{Gross_1968}. Equations (\ref{M_g_M_gamma}) and (\ref{H constant}) lead to the same proportionality between spectra found in Refs. \cite{Nikishov_1989,Nikishov_2010} but a relation like Eq. (\ref{M_g_M_gamma}) is not expected to hold for a higher number of incoming photons \cite{Ahmadiniaz_2019_a} because the arguments based on conservation laws \cite{Goebel_1980} cease to apply. Thus, an investigation of this property in the context of strong-field QED, where the effects of an electromagnetic background are taken into account exactly, is certainly relevant and timely.
	
	In the present paper we show both classically and quantum mechanically that the electromagnetic and the gravitational radiation amplitudes in an arbitrary plane-wave background field are proportional to each other and that the proportionality constant is the same in both cases and equal to $H$. Classically, this is achieved by introducing a concept of radiation amplitude in analogy with the quantum one. Quantum mechanically we work within strong-field QED in the Furry picture, which allows to take into account exactly the effects of the plane wave into the electron dynamics \cite{Landau_b_4_1982} \corr{(see also Ref. \cite{Galtsov_1975} for a similar computation in a circularly-polarized monochromatic plane wave)}. It is remarkable that the proportionality relies only on the symmetries of the background plane wave and on the semiclassical nature of the quantum dynamics of electrons in a plane wave.
	
	\section{Proportionality between classical gravitational and electromagnetic radiation amplitudes}

	An electron in the presence of an intense electromagnetic plane wave radiates both light and gravitational waves. The electron is characterized by an initial four-momentum $p$, whereas the plane wave is described by the four-vector potential $A^{\mu}_B(\phi)$, where $\phi=n_B\cdot x$, with $n^{\mu}_B=(1,\bm{n}_B)$,  $\bm{n}_B$ being the unit vector along the propagation direction of the plane wave. We assume that $A^{\mu}_B(\phi)$ satisfies the Lorenz-gauge condition and $A^0_B(\phi)=0$, and that $\lim_{\phi\to\pm \infty}A^\mu_B(\phi)=0$. Below, we will use the light-cone notation by setting $\tilde{n}_B^\mu = \frac{1}{2}(1, - \bm{n}_B)$ and $a_{1,2}^\mu = (0, \bm{a}_{1,2})$ such that $\bm{a}_i\cdot \bm{n}_B = 0$ and $\bm{a}_i \cdot \bm{a}_j = \delta_{ij}$. The quantities $\{n_B^\mu ,\tilde{n}_B^\mu, a_{1,2}^\mu \}$ form a basis of the four-vector space such that an arbitrary four-vector $v^\mu$ can be written as $v^\mu = v_-\tilde{n}^\mu_B + v_+ n_B^\mu + v_\perp ^\mu$, where $v_- = v \cdot n_B$, $v_+ = v \cdot \tilde{n}_B $, and $v_{\perp}^\mu  = -\sum_{i=1}^2(v \cdot a_i ) a_i^\mu$.
	
	According to Einstein's equations, any form of matter or energy is a source of gravity. Due to their remarkably small amplitudes in most situations, we will consider here linear gravitational waves in the first-order weak-field approximation $\mathcal{O}(\kappa)$, where the metric is expanded as $g_{\mu\nu} = \eta_{\mu\nu} + \kappa h_{\mu\nu}$ \cite{Weinberg_1972, Maggiore_2007}, with $\eta_{\mu\nu}=\text{diag}(+1,-1,-1,-1)$. This implies that the energy-momentum brought by the gravitational wave itself is not included in the total energy-momentum tensor $T^{\mu\nu}$, which is taken as the source of the gravitational field. Thus, the sources of gravitational waves in the system under consideration are the particle $P$, the background field $A^{\mu}_B$ and the field radiated by the charge $A^{\mu}_R$. Moreover, the electromagnetic stress tensor $T_{EM}^{\mu\nu}=F^{\mu \alpha}F_{\alpha}^{ \-\ \nu}  + \frac{1}{4}\eta^{\mu\nu}F^{\alpha \beta}F_{\alpha \beta}$ is quadratic in the field $F^{\mu\nu} = F_R^{\mu\nu} +F_B^{\mu\nu}$, where $F^{\mu\nu}_{R,B} = \partial^{[\mu}A_{R,B}^{\nu]}$, with $a^{[\mu}b^{\nu]} = a^\mu b^\nu - a^\nu b^\mu$, and therefore a mixed term arises involving background and radiation field. Consequently, on the whole we have $T^{\mu\nu} = T_P^{\mu\nu} + T_R^{\mu\nu} + T_B^{\mu\nu} + T_{RB}^{\mu\nu}$, where
\begin{align}
					T_P^{\mu\nu}&=m\int d\tau u^{\mu}(\tau)u^{\nu}(\tau)\delta^{(4)}(x-x(\tau)),\\
					T_{B,R}^{\mu\nu} &= F_{B,R}^{\mu \alpha}F_{ B,R,\alpha}^{ \quad \nu} 
					+ \frac{1}{4} \eta^{\mu\nu} F_{B,R}^{\alpha \beta}F_{B,R,\alpha \beta},\\ 
					T_{RB}^{\mu\nu} &= 
					F_{R}^{\mu \alpha}F_{ B,\alpha}^{ \quad \nu} 
					+ F_{B}^{\mu \alpha}F_{ R, \alpha}^{ \quad  \nu} 
					+ \frac{1}{2}\eta^{\mu\nu}F_{R}^{\alpha \beta}F_{B,\alpha \beta},
				\end{align}
with $x^{\mu}(\tau)$ being the electron trajectory parametrized via the proper time $\tau$ and $u^{\mu}(\tau)=dx^{\mu}/d\tau$.
	
	Now, we are interested in a regime where the background plane wave can be intense, in the sense that the classical nonlinearity parameter $\xi = |e| E_B /m\omega_B $ can be larger than unity \cite{Di_Piazza_2012,Gonoskov_2022,Fedotov_2022}. Here, $E_B$ and $\omega_B$ are the peak value of the electric field of the wave and its typical angular frequency, respectively. We work in a parameter range, where radiation-reaction effects can be neglected. Classically, this implies that the parameter $\alpha\chi\xi\Phi$ is much smaller than unity \cite{Di_Piazza_2008_a}, where $\chi=(p_-/m)(E_B/E_{cr})$, with $p^{\mu}$ being the initial four-momentum of the electron and $E_{cr}=m^2/|e|$ the critical field of QED \cite{Fradkin_B_1991,Di_Piazza_2012,Gonoskov_2022,Fedotov_2022}, and where $\Phi$ is the total phase duration of the plane-wave field. Quantum mechanically, this implies that multiple photon emissions and radiative corrections are negligible, which is the case if $\alpha\xi\Phi\ll 1$ at $\chi\lesssim 1$ \cite{Di_Piazza_2010,Podszus_2021}. Moreover, neglecting radiation-reaction effects means classically that the conservation of the total energy-momentum tensor has to be equivalent to the electron dynamics being described by the Lorentz equation and this implies that the energy-momentum tensor $T_{R}^{\mu\nu}$ of the electromagnetic field produced by the electron can be ignored in $T^{\mu\nu}$. It should be stressed that if $T_{R}^{\mu\nu}$ is taken into account, analytical problems arise because of its divergence on the electron trajectory. This divergence is not avoidable unless one introduces a finite size model for the electron. In this way, we have that $T^{\mu\nu} = T_P^{\mu\nu} +  T_B^{\mu\nu} + T_{RB}^{\mu\nu}$ (see also Refs. \cite{Nikishov_1989,Nikishov_2010}). 
	
	At this point, one would expect the largest electromagnetic contribution to the gravitational field to come from the background term $T_B^{\mu\nu}=-A_B^{\prime\,2}n_B^{\mu}n_B^{\nu}$, where $A_B^{\prime\,\mu}=dA^{\mu}_B/d\phi$. However, since $T_B^{\mu\nu}$ depends only on $\phi$, its Fourier transform $T_B^{\mu\nu}(k) =(2 \pi)^3\delta(k_-) \delta^{(2)}(\bm{k}_\perp)\rho(k_+) n_B^\mu n_B^\nu$, with $\rho(k_+)=-\int d\phi \exp(ik_+\phi)A_B^{\prime\,2}(\phi)$ does not vanish only for $k^{\mu}\propto k_B^{\mu}$, with $k_B^{\mu}=\omega_Bn_B^{\mu}$, such that, due to gauge invariance, $T_B^{\mu\nu}(k)\varepsilon^*_{\mu}\varepsilon^*_{\nu}=0$ . Therefore $T_B^{\mu\nu}$ does not contribute to the gravitational radiation (see also below) \cite{Landau_b_2_1975}. We then conclude that \cite{Nikishov_1989, Nikishov_2010}
	\begin{equation}
		\label{total EM tensor}
		T^{\mu\nu} = T_P^{\mu\nu}+ T_{RB}^{\mu\nu}.
	\end{equation}
	
	Classically, we can introduce the amplitudes $S^\gamma_c(k)$ and $S^g_c(k)$ of the emission of electromagnetic and gravitational radiation, respectively, in such a way that they coincide with the first-order quantum counterparts: 
	\begin{align}
		S^\gamma_c(k) 	&=  - ie J_P^\mu(k) \varepsilon^*_\mu,\\
		S_c^g(k) &= i \frac{\kappa}{2} T^{ \mu\nu}(k)\varepsilon^*_\mu \varepsilon^*_\nu,
	\end{align}
	where $J_P^{\mu}(k)$ is the Fourier transform of the four-current $J_P^\mu(x)= \delta^{(3)}(\bm{x}- \bm{x}(t))dx^\mu(t)/dt$ of the electron moving along the trajectory $x^{\mu}(t)$. As we have mentioned in the introduction, $\varepsilon^\mu$ ($\varepsilon^{\mu}\varepsilon^{\nu}$) is the helicity polarization four-vector (tensor) of the electromagnetic (gravitational) wave such that $\varepsilon \cdot \varepsilon = k \cdot \varepsilon = 0$ \cite{Gross_1968}. By squaring the above amplitudes and summing over the polarizations, one finds the corresponding energy emission spectra (see Refs. \cite{Weinberg_1972,Jackson_B_1975,Nikishov_2010}):
	\begin{align}\label{spectra definitions_1}
		\begin{split}
			\frac{ d \mathcal{E}_\gamma}{ d \bm{k}} & = \inv{16 \pi^3} \sum_{\text{pol.}} S_c^{\gamma *}(k)\, S^\gamma_c(k) \\
			&= - \frac{e^2}{16 \pi^3}\,  J_{P,\mu}^*(k)J_P^\mu(k),
		\end{split}\\
		\label{spectra definitions_2}
		\begin{split}
			\frac{d \mathcal{E}_g}{d\bm{k}} & = \frac{1}{16 \pi^3}  \sum_{\text{pol.}}  S_c^{g *}(k)S^g_c(k)
			\\ &
			= \frac{\kappa^2}{64 \pi ^3}\left[
			T^{\mu\nu}(k) T^*_{\mu\nu}(k) -\frac{1}{2}T^{\mu}_{\-\ \mu}(k)T^{*\nu}_{\-\ \nu}(k) 
			\right],
		\end{split}
	\end{align}
	where the completeness relations 
	$\sum_{\text{pol.}} \varepsilon_{\mu}\varepsilon^*_\nu \Rightarrow  - \eta_{\mu\nu}\quad $ \cite{Landau_b_4_1982} and $
	\sum_{\text{pol.}} \varepsilon_{\mu \nu }\varepsilon^*_{\alpha \beta} \Rightarrow  \left( \eta_{\mu\alpha} \eta_{\nu \beta} + \eta_{\mu\beta} \eta_{\nu\alpha} - \eta_{\mu\nu} \eta_{\alpha\beta} \right)/2$ were used \cite{Veltman_1975}.
	
	In order to show the proportionality between $S_c^g(k)$ and $S_c^\gamma(k)$, we first consider the mixed energy-momentum tensor $T^{\mu\nu}_{RB}(k)$ in Fourier space, which can be written as
	\begin{equation}
		\begin{split}
			T^{\mu \nu}_{RB}(k) = & \-\
			\int\frac{d^4q}{(2\pi)^4}\bigg[
			\inv{2}F_B^{\alpha \beta}(k-q)F_{R,\alpha \beta}(q)\eta^{\mu\nu}
			\\ & 
			+F_B^{\mu\alpha}(k-q)F_{R,\alpha}^{ \quad   \nu}(q) +F_B^{\nu\alpha}(k-q)F_{R,\alpha}^{\quad  \mu}(q)
			\bigg].
		\end{split}
	\end{equation}
	
	By employing the retarded solution of the wave equation $\square A^\mu _R = eJ_P^\mu$, the field tensor $F_R^{\mu\nu}(q)$ in momentum space is given by
	\begin{equation}\label{Max. tensor R}
		\begin{split}
			F_R^{\mu\nu}(q) = \int d^4x \, e^{iq\cdot x}  \partial^{[\mu}A_R^{\nu]}(x) 
			= 	 \frac{ie}{q^2+i\epsilon q^0} q^{[\mu}J_P^{\nu]}(q),
		\end{split} 
	\end{equation}
	and, since $\varepsilon\cdot\varepsilon=0$, we obtain
	\begin{equation}\label{Amplitude 1}
		\begin{split}
			S_c^g(k)  & =  \frac{\kappa}{2} \bigg[i T^{\mu\nu}_P(k)
			\\ &\quad
			\left. -2e \int\frac{d^4q}{(2\pi)^4}
			\frac{F_B^{\mu\alpha}(k-q)q_{[\alpha}J^{\nu]}_P(q)}{q^2+i\epsilon q^0}  
			\right] \varepsilon^*_\mu  \varepsilon^*_\nu .
		\end{split}	
	\end{equation}
	The energy-momentum tensor of the electron in Fourier space is \cite{Weinberg_1972}
	\begin{equation}\label{T particle}
		T_P^{\mu\nu} (k)= 
		( \pi^\mu J^\nu_P ) (k) =  
		\int\frac{d^4 q}{(2 \pi)^4} \pi^\mu(k-q)J^\nu_P(q) ,
	\end{equation}
	where $\pi^\mu (\phi)$ is the electron four-momentum in the plane wave $A_B^{\mu}$, with the initial condition $\lim_{\phi\to-\infty}\pi^\mu(\phi) = p^\mu$. Now we observe that
	\begin{equation} \label{Max. tensor B}
		F_B^{\mu\alpha}(q) = 
		(2 \pi)^3\delta(q_-) \delta^{(2)}(\bm{q}_\perp) F_B^{\mu\alpha}(q_+),
	\end{equation}
	and that in the chosen gauge it is $\pi^{\mu}_\perp (\phi) = p^{\mu}_\perp -e A^{\mu}_{B,\perp}(\phi)$, such that
	\begin{equation}\label{Max tensor B momentum}
		F_B^{\mu\nu}(q_+) = 
		\frac{i q_+}{e} [2\pi \delta (q_+)  p_{\perp}^{[\mu} n_B^{\nu]} 
		-\pi_\perp^{[\mu}(q_+)n_B^{\nu]}].
	\end{equation}
	It is important to notice that one cannot use the identity $q_+\delta(q_+)=0$ here because of the term $q^2+i\epsilon q^0$ in the denominator of Eq. (\ref{Amplitude 1}). In general, the product involving the delta-function is not associative and, by replacing Eqs. (\ref{Max. tensor B}) and (\ref{Max tensor B momentum}) in Eq. (\ref{Amplitude 1}), one finds that
	\begin{equation}\label{Amplitude Fourier classic}
		\begin{split}
			& S^g_c(k)   = 
			\frac{i \kappa}{2 k_-} 
			\left( k\cdot n_B  p_{\perp} \cdot \varepsilon^*  - k \cdot p_{\perp}  n_B \cdot \varepsilon^*   \right) J_P(k) \cdot \varepsilon^*  
			\\ &   \qquad  \qquad +
			\frac{\kappa}{2} \bigg[
			i T^{\mu\nu}_P(k) -
			\frac{i}{k_-}
			k_\alpha \bigg(\pi_\perp^{[\mu}n_B^{\alpha]}J_P^{\nu}\bigg)(k) 
			\\ & \quad  \quad  \qquad \qquad \qquad \qquad 
			+ \frac{e}{k_-} \bigg( F_B^{\mu \alpha}J_{P,\alpha} \bigg) (k) n_B^\nu
			\bigg]
			\varepsilon^*_\mu  \varepsilon^*_\nu,
		\end{split}
	\end{equation}
	where we have used the fact that in Eq. (\ref{Amplitude 1}) on can replace $q^{\mu}-k^{\mu}=(q_+-k_+)n_B^{\mu}$. 
	
	At this point one can exploit the equation of motions in Fourier space 
	\begin{equation} \label{Eq of motion Fourier classic}
		\partial_\nu T_P^{\mu\nu} =  
		eF_B^{\mu \alpha} 
		J_{P, \alpha}
		\; \Rightarrow \;
		e\left(F_B^{\mu \alpha} 
		J_{P,\alpha} \right)(k)	= -ik_\nu T_P^{\mu\nu} (k) 
	\end{equation}
	and use Eq. (\ref{T particle}) to obtain an expression of $S^g_c(k)$ depending only on the electron four-current and four-momentum:
	\begin{equation}
		\begin{split}
			S^g_c(k) &  = \frac{i\kappa}{2 k_-} 
			\left( k\cdot n_B  p_{\perp} \cdot \varepsilon^*  - k \cdot p_{\perp}  n_B \cdot \varepsilon^*   \right) J_P(k) \cdot \varepsilon^*  
			\\ &\quad +
			\frac{i\kappa}{2  k_-} \bigg[
			- k_-   \pi_\perp^{\mu} J_P^{\nu} +
			k \cdot  \pi_\perp  n_B^{\mu}J_P^{\nu} 
			\\ & \qquad \qquad  \qquad 
			-  k \cdot \pi  n_B^\mu   J_P^\nu 
			+ k_-   \pi^\mu  J_P^\nu 
			\bigg](k)
			\varepsilon^*_\mu  \varepsilon^*_\nu .
		\end{split}
	\end{equation}
	Finally, by observing that $k\cdot \pi = k_- \pi_+ + k_+ \pi_- + k_\perp \cdot \pi_\perp$ and that $\pi_- = p_-$, the following proportionality is found 
	\begin{equation}
		\label{Prop_c}
		S^g_c(k) = - \frac{\kappa}{2 e}
		\left(
		\frac{p \cdot  \varepsilon^* k \cdot k_B - k \cdot p k_B \cdot \varepsilon^*  }{k \cdot k_B}
		\right) S^\gamma_c(k).
	\end{equation}
	
	On the one hand, the proportionality constant in Eq. (\ref{Prop_c}) coincides with that in Eq. (\ref{H constant}), as it can be easily verified by using the relations $p^{\mu}_i+k^{\mu}_i = p^{\mu}_f+k^{\mu}_f$ and $k_f \cdot \varepsilon_f^* = 0$ and by identifying $p^{\mu}_i=p^{\mu}$, $k^{\mu}_i=k^{\mu}_B$, $k_f^{\mu}=k^{\mu}$, and $\varepsilon_f^{\mu}=\varepsilon^{\mu}$. Thus, we indicate it also as $H$, i.e., $S^g_c(k) = HS^\gamma_c(k)$. On the other hand,	Eq. (\ref{Prop_c}) generalizes the results in Refs. \cite{Nikishov_1989,Nikishov_2010} as here the proportionality is shown to exist already at the level of the amplitudes and for an arbitrary plane wave, whereas in Refs. \cite{Nikishov_1989,Nikishov_2010}  the proportionality was found in the energy spectra and for a monochromatic plane wave. Indeed, finding the gravitational energy spectrum is straightforward [see Eqs. (\ref{spectra definitions_1}) and (\ref{spectra definitions_2})]:
	\begin{equation}
		\der{\mathcal{E}_g}{\bm{k}}  =  
		-\inv{2}\left(\pder{H}{\varepsilon^*_\mu}\right)^2 \der{\mathcal{E}_\gamma}{\bm{k}},
	\end{equation}
	where
	\begin{equation}
\label{H_2}
		-\inv{2}\left(\pder{H}{\varepsilon^*_\mu}\right)^2 = 
		-\frac{4 \pi G}{e^2}\left(m^2   - 2 \frac{p \cdot k p \cdot k_B}{k \cdot k_B}\right)
	\end{equation}
	in agreement with Ref. \cite{Nikishov_2010}, once one takes into account that there the calculations are carried out for a monochromatic plane wave in the average rest frame of the electron and that the authors of Ref. \cite{Nikishov_2010} use the opposite Minkowski metric tensor as compared to ours. It is instructive to show the agreement explicitly. Since the proportionality constant is a Lorentz-invariant quantity, we can assume without loss of generality that in the laboratory frame the electron is initially at rest. Below, the subscript $L$ ($R$) indicates quantities in the laboratory (average rest) frame\footnote{Note that the index $R$ has this meaning only until the end of the section and it should not be confused with the index $R$ in rest of the paper.}. The latter has a relative velocity as compared to the former given by $\bm{\beta}_d = \beta_d \bm{n}_B$, with $\beta_d$ defined via the relation
	\begin{equation}
	\langle \bm{\pi}_R(\phi) \cdot \bm{ n}_B \rangle = \gamma(\beta_d)
	\big[
	\langle \bm{\pi}_L(\phi) \cdot \bm{ n}_B \rangle  - \beta_d \langle \varepsilon_L(\phi) \rangle      
	\big]
	  = 0,
	\end{equation}
	where the averages are taken over a plane wave period. In the average rest frame the electron energy corresponds to the so-called effective mass $m_* = m \gamma_*$ \cite{Nikishov_2010}, such that $\gamma_*  = \sqrt{(1 + \beta_d)/(1-\beta_d)}$ describes the relativistic Doppler effect: $\omega_{B,L} = \gamma_* \omega_{B,R}$ (recall that $\bm{\beta}_d$ is along the propagation direction of the laser). In the laboratory frame it is $p_{L,-}=m$, such that
	\begin{equation}
	    m_* = \frac{p_{L,-}\omega_{B,L}}{\omega_{B,R}} = \frac{k_{B,L} \cdot p_L }{\omega_{B,R}}.
	    \end{equation}
	    Thus, since the quantity $k_B \cdot p$ is Lorentz invariant, one finds $m_* = p_{R,-}$. 
	The proportionality constant introduced in Ref. \cite{Nikishov_2010} [see Eqs. (1.5), (1.6), and (2.23) there] can then be written as
	\begin{equation}
		C = \frac{4\pi G}{e^2}\frac{p_{R,-}^2 \bm{k}_{R,\perp}^2}{k_{R,-}^2}.
	\end{equation}
	Finally, the equivalence of the two expressions of the proportionality constant is obtained by observing that in the average rest frame the electron is initially counterpropagating with respect to the plane wave so that $\bm{p}_{R,\perp} = \bm{0}$ and therefore [see Eq. (\ref{H_2})]
	\begin{equation}
		\begin{split}
			&-\frac{4 \pi G}{e^2}\left(m^2   - 2 \frac{p_R \cdot k_R \,  p_R \cdot k_{B,R}}{k_R \cdot k_{B,R}}\right)=  \frac{4\pi G}{e^2}\frac{p_{R,-}^2 \bm{k}_{R,\perp}^2}{k_{R,-}^2}.
		\end{split}
	\end{equation}

	\section{Graviton photoproduction at tree level in strong-field QED}
		
\begin{figure*}[t] 
		\centering
		\includegraphics[width=1\linewidth]{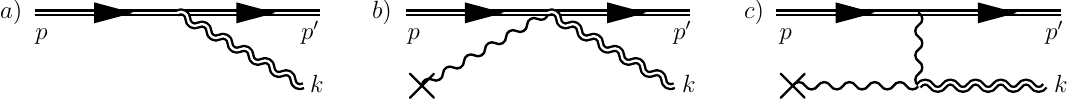} 
		\caption{Feynman diagrams contributing to the graviton photoproduction at the order $\mathcal{O}(\kappa)$ in the presence of a strong electromagnetic plane-wave background field. The double oscillating lines represent the graviton, while the double fermion lines correspond to Volkov states, and the oscillating lines with the cross represent the background field source.}
		\label{fig:feynman-diagrams}
	\end{figure*}

	Now, we pass to the quantum case. By linearizing the Einstein-Hilbert action \cite{Weinberg_1972,Landau_b_2_1975} one obtains a field theory for the graviton  $h_{\mu\nu}$ describing a spin-2 massless particle \cite{Maggiore_2007, Weinberg_1972}. Working in the de Donder (or harmonic) gauge, the Lagrangian density of the field $h_{\mu\nu}$ coupled to a generic, conserved energy-momentum tensor $T^{\mu\nu}$ is given by \cite{Maggiore_2007}
	\begin{equation} \label{ L graviton plus matter }
		\mathcal{L}_g  =   -\inv{2}
		\partial_\alpha h_{\mu\nu}\partial^\alpha h^{\mu\nu}
		+ \inv{4} \partial^\mu  h   \partial_\mu h 
		+ \frac{\kappa}{2} h_{\mu\nu} T^{\mu\nu},
	\end{equation}
	where $h = h^\mu_{\-\ \mu}$. The electromagnetic sector is described by the strong-field QED Lagrangian density
	\begin{equation}
		\mathcal{L}_{\gamma} = - \frac{1}{4}F_Q^{\mu\nu}F_{Q,\mu\nu} +\bar{\psi}\left( i \slashed{\partial} - e \slashed{A}_B - m \right) \psi -eA_Q^\mu J_{D,\mu}, 
	\end{equation}
	where $J^\mu_D = \bar{\psi}\gamma^\mu \psi$ is the Dirac four-current and $A_Q^\mu$ is the photon field ($F_Q^{\mu\nu}=\partial^{\mu}A_Q^{\nu}-\partial^{\nu}A_Q^{\mu}$). We assume to work within the Furry picture \cite{Furry_1951}, where the Dirac field is quantized in the presence of the plane-wave field $A_B^{\mu}$. Thus, for instance, the positive-energy state for an electron with four-momentum $p^{\mu}$ outside the plane wave is given by the Volkov spinor \cite{Volkov_1935,Landau_b_4_1982}
	\begin{equation}
		\psi_p(x) = e^{ i S_p(x)} \left[  1 + \frac{e \slashed{n}_B\slashed{A}_B(\phi)}{2 n_B \cdot p}   \right] u_p,
	\end{equation}
	where $S_p$ is the classical action of an electron in a plane wave \cite{Landau_b_2_1975, Di_Piazza_2012}: $- \partial^\mu S_p(x) = \pi^\mu_p(\phi) + e A_B^\mu (\phi)$ and $u_p$ is the free positive-energy spinor (for notational simplicity, the spin quantum number is not indicated). The energy momentum tensor coupled to the field $h_{\mu\nu}$ in Eq. (\ref{ L graviton plus matter }) is $T^{\mu\nu} = T_D^{\mu\nu} + T_Q^{\mu\nu} + T_{QB}^{\mu\nu} + T_B^{\mu\nu}$,
	where 
	\begin{equation}
		\begin{split}
			& T_D^{\mu\nu}  =  
			\bar{\psi}\left[
			\frac{i}{4}  \gamma^{\{\mu}  \overleftrightarrow{\partial}^{\nu \}} - 
			\frac{e}{2} \gamma^{\{\mu} A_B^{\nu \}} 
			\right. \\  & \left. \qquad \qquad \qquad \qquad
			- \eta^{\mu\nu} \left(  \frac{i}{2} \overleftrightarrow{\slashed{\partial}}- e \slashed{A}_B - m     \right)
			\right]\psi,
			\\ & 
			T_{Q}^{\mu\nu} = 
			F_Q^{\mu \alpha}F_{ Q,\alpha}^{ \quad \nu} 
			+ \frac{1}{4} \eta^{\mu\nu} F_Q^{\alpha \beta}F_{Q,\alpha \beta} ,
			\\ & 
			T_{QB}^{\mu\nu} = 
			F_Q^{\mu \alpha}F_{ B,\alpha}^{ \quad \nu} 
			+ F_B^{\mu \alpha}F_{ Q, \alpha}^{ \quad  \nu} 
			+ \frac{1}{2}\eta^{\mu\nu}F_Q^{\alpha \beta}F_{B,\alpha \beta},
		\end{split} 
	\end{equation}
	with $a^{\{\mu}b^{\nu\}} = a^\mu b^\nu + a^\nu b^\mu$ and $\bar{\psi} \overleftrightarrow{\partial}^{\nu}\psi = \bar{\psi} \overrightarrow{\partial}^{\nu}\psi  - \bar{\psi} \overleftarrow{\partial}^{\nu}\psi $.
	The $S$-matrix transition amplitude of the graviton photoproduction by an electron driven by an intense plane wave ($\xi \gtrsim 1$) is given by $S^g_{fi} =\langle p';k,\varepsilon_{\mu\nu} |S| p\rangle$, where the initial and the final electron states are Volkov states. The process $e \rightarrow e + g $ here is allowed because the background plane wave supplies the otherwise missing energy-momentum.  The $S$-matrix is defined as $S=\mathcal{T}\exp\left[i\int d^4x\,\mathcal{L}_i(x)\right]$, where $\mathcal{T}$ is the time-ordering operator and
	\begin{equation}
		\mathcal{L}_i=\frac{\kappa}{2} h_{\mu\nu} T^{\mu\nu}-eA_Q^\mu J_{D,\mu}.
	\end{equation}
	At the first order in $\kappa$ the process is described by the Feynman diagrams in Fig. \ref{fig:feynman-diagrams} 
	and the corresponding amplitude is given by
	\begin{equation}
		\begin{split}
			S^g_{fi} =  & 
			i \frac{\kappa}{2} \mathcal{T}
			\langle p' |
			\bigg\{
			\int d^4 x\, e^{ik \cdot x}  \varepsilon^*_\mu \varepsilon^*_\nu
			\bigg[
			T_D^{\mu \nu}(x) 
			\\&  \qquad \quad 
			-ie T_{QB}^{\mu\nu}(x)	\int d^4y \,  J_{D,\alpha}(y) A_Q^\alpha(y) 
			\bigg]
			\bigg\}
			| p\rangle.
		\end{split}
	\end{equation}
	As in the classical case $T_B^{\mu\nu}$ does not contribute because of gauge invariance. Also, it is easily seen that this matrix element has exactly the same form of the classical amplitude Eq. (\ref{Amplitude 1}), the only differences being the photon propagator which now has to follow the Feynman prescription and the spinorial nature of the particle:
	\begin{equation}\label{Amplitude quantum}
		\begin{split}
			& S^g_{fi}  
			= \frac{\kappa}{2} \Bigg[ i T^{\mu\nu}_V(k)  \\ & \left.  
			- 2e\int\frac{d^4q}{(2\pi)^4 }
			\frac{F_B^{\mu\alpha}(k-q) q_{[\alpha} J^{\nu]}_V(q) }{q^2 + i\epsilon }
			\right] \varepsilon^*_\mu  \varepsilon^*_\nu,
		\end{split}	
	\end{equation}
	where
	\begin{align}
		J_V^\mu &= \langle p'|J_D^{\mu} | p\rangle  = \bar{\psi}_{p'}\gamma^\mu \psi_p,\\
		T^{\mu\nu}_V &= \langle p'|T^{\mu\nu}_D | p\rangle=\bar{\psi}_{p'}\left(
		\frac{i}{4}  \gamma^{\{\mu}  \overleftrightarrow{\partial}^{\nu \}} -
		\frac{e}{2} \gamma^{\{\mu} A_B^{\nu \}} \right)\psi_p
	\end{align}
	are the matrix elements of the corresponding operators between Volkov states. \corr{Although the structure of the amplitude is similar to the classical one in Eq. (\ref{Amplitude 1}), we stress the fact that quantum effects like spin effects and the recoil on the electron are taken into account in Eq. (\ref{Amplitude quantum})}. Now, the considerations about the background field corresponding to Eqs. (\ref{Max. tensor B}) - (\ref{Max tensor B momentum}) clearly remain valid here. Thus, it is easily seen that the retarded and the Feynman prescriptions lead to the same result because the minus and the perpendicular components of $q^{\mu}$ are fixed by the conservation laws and the remaining term $k_+-q_+$ in the denominator is compensated as in the classical case. Consequently, one can derive the analogous of Eq. (\ref{Amplitude Fourier classic}), which now reads
	\begin{equation}\label{Amplitude Fourier quantum}
		\begin{split}
			& S^g_{fi}  = 
			i \frac{\kappa}{2 k_-} 
			\left( k\cdot n_B p_{\perp} \cdot \varepsilon^*  - k \cdot p_{\perp}   n_B \cdot \varepsilon^*   \right) J_V (k) \cdot \varepsilon^* 
			\\ & \quad +
			\frac{\kappa}{2} \bigg[
			i  T^{\mu\nu}_V (k) -
			\frac{i}{k_-}
			k_\alpha \bigg(\pi_\perp^{[\mu}n_B^{\alpha]} J^{\nu}_V  \bigg)(k) \-\ 
			\\ & \qquad \qquad \qquad \qquad  
			+ \frac{e}{k_-} \bigg( F_B^{\mu \alpha}J_{V,\alpha}   \bigg) (k)  n_B^\nu
			\bigg]
			\varepsilon^*_\mu  \varepsilon^*_\nu .
		\end{split}
	\end{equation}
	Moreover, the equations of motion for the matrix elements $T^{\mu\nu}_V$ and $J_V^\mu$ are the same as for the classical quantities [see Eq. (\ref{Eq of motion Fourier classic})]:
	\begin{equation} \label{Eq of motion Fourier quantum}
		e\left(F_B^{\mu \alpha} 
		J_{V, \alpha} \right)(k)= -ik_\nu T_V^{\mu\nu} (k).
	\end{equation}
	However, due to the spinorial structure of Volkov states, the quantity $T^{\mu\nu}_V$ cannot be put in the classical form as in Eq. (\ref{T particle}) but it can rather be written as
	\begin{equation}
		T^{\mu\nu}_V (k)\varepsilon^*_\mu \varepsilon^*_\nu  =
		\left(
		J_V^\mu  \pi^\nu_p 	+ i  \frac{e }{4p_-}	\bar{\psi}_{p'}
		\gamma^\mu \slashed{F}_B
		\psi_p  n_B^\nu 
		\right)\varepsilon^*_\mu \varepsilon^*_\nu,
	\end{equation} 
	where $\slashed{F}_B = F^{\alpha \beta}_B \gamma_\alpha \gamma_\beta$. Interestingly, the spin terms do cancel in the combination $iT^{\mu\nu}_V - (i/k_-)k_\alpha T^{\mu\alpha}_V n_B^\nu$ in Eq. (\ref{Amplitude Fourier quantum}). Thus, in conclusion, the same result as in the classical treatment is found 	
	\begin{equation}
		S^g_{fi} = - \frac{\kappa}{2e}
		\left(
		\frac{p \cdot  \varepsilon^*  n \cdot n_B - n \cdot p n_B \cdot \varepsilon^*  }{n \cdot n_B}
		\right) S^\gamma_{fi}, 
	\end{equation}
	where $S^\gamma_{fi} = - ie J_V(k)\cdot \varepsilon^* $ is the matrix element of nonlinear Compton scattering (see, e.g., Ref. \cite{Landau_b_4_1982}).
	
	A comment is in order, which pertains to both the classical and the quantum regime. The proportionality constant $H$ diverges as $1/\theta$ in the limit where the angle $\theta$ between the graviton and the plane-wave photons tends to zero (collinear emission). Since in the same limit, the Compton-scattering probability tends to zero linearly \cite{Landau_b_4_1982}, one concludes that the graviton-emission probability diverges logarithmically \cite{Bjerrum-Bohr_2015, Nikishov_2010}. The same is true for the classical gravitational energy spectrum \cite{Nikishov_2010}. For small scattering angles the dominant contribution comes from the interaction between the particle field and the background. Classically this can be seen from the trend of the formation length which grows with the collinearity because the radiated electromagnetic field and the background interact for a longer and longer time before the gravitational conversion takes place \cite{Nikishov_2010}. Since the mixed electromagnetic energy-momentum tensor grows with the formation length, the emission probability increases. Quantum mechanically this corresponds to the dominance of the $t$-channel diagram in Fig. \ref{fig:feynman-diagrams} (c) in this limit. Indeed, it is $t  = (p-p')^2 \propto k \cdot k_B $ and when this goes to zero an infinite contribution arises from the photon propagator. It is worth noting that this diagram is dominant also in the non-relativistic range where the photon recoil is negligible and $p \rightarrow p'$ \cite{DeLogi_1977qe}.
	
	\section{Conclusions}

	To summarize, we have shown both classically and quantum mechanically that the amplitudes of graviton and photon emission by an electron in an arbitrary plane wave are proportional to each other. Although the electron dynamics is highly nonlinear in the plane wave and quantum effects are large, the proportionality constant is classical and it does not depend on the plane-wave intensity. At the fundamental quantum level, by combining strong-field QED and quantum gravity, our proof shows that the proportionality relies only on the symmetries of the plane wave and the semiclassical nature of the motion of a quantum particle in a plane wave background.

\end{document}